\documentclass[11pt]{extarticle}
\usepackage[paperheight=12in,paperwidth=8.75in]{geometry}
\usepackage{setspace}

\usepackage{setspace}
\usepackage[T1]{fontenc}
\usepackage{times}
\usepackage{palatino}
\usepackage{lmodern}
\usepackage[latin1]{inputenc}
\usepackage{epsfig}
\usepackage[english]{babel}
\usepackage{color}
\usepackage{graphicx}
\usepackage{dcolumn}
\usepackage{amsmath,amssymb,amsfonts}
\usepackage[all]{xy}
\usepackage{cite}
\usepackage[dvipsnames]{xcolor}
\usepackage{physics}
\usepackage{mathrsfs}
\usepackage{authblk}
\usepackage{geometry}
\usepackage{abstract}
\usepackage[english]{babel}
\usepackage[T1]{fontenc}
\usepackage[colorlinks,urlcolor=blue,citecolor=blue]{hyperref}

\begin{document}
\numberwithin{equation}{section}
\newcommand{\boxedeqn}[1]{%
  \[\fbox{%
      \addtolength{\linewidth}{-2\fboxsep}%
      \addtolength{\linewidth}{-2\fboxrule}%
      \begin{minipage}{\linewidth}%
      \begin{equation}#1\end{equation}%
      \end{minipage}%
    }\]%
}


\newsavebox{\fmbox}
\newenvironment{fmpage}[1]
     {\begin{lrbox}{\fmbox}\begin{minipage}{#1}}
     {\end{minipage}\end{lrbox}\fbox{\usebox{\fmbox}}}

\raggedbottom
\onecolumn

\begin{center}
{\Large \bf
Quadratic symmetry algebra and spectrum of the 3D nondegenerate quantum superintegrable system\\
}
\vspace{6mm}
{ Mohasena Ahamed$^{a}$ and Md Fazlul Hoque$^{b,c}$
}
\\[6mm]
\noindent ${}^{a}${\em 
Daffodil International University, Department of General Education Development, Ashulia, Bangladesh}\\[3mm]
\noindent ${}^{b}${\em 
Pabna University of Science and Technology, Department of Mathematics, Pabna 6600, Bangladesh}\\[3mm]
\noindent ${}^c${\em
Czech Technical University in Prague, Faculty of Nuclear Sciences and Physical Engineering,
 Department of Physics, B\v{r}ehov\'a 7, 115 19 Prague 1, Czech Republic}\\ [1mm]
\vspace{4mm}
{\footnotesize Email: mohasena.math@gmail.com, fazlulmath@pust.ac.bd}
\end{center}
\vskip 1cm

\vskip 1cm
\begin{abstract}
\noindent 
In this paper, we present the quadratic associative symmetry algebra of the 3D nondegenerate maximally quantum superintegrable system. This is the complete symmetry algebra of the system. It is demonstrated that the symmetry algebra contains suitable quadratic subalgebras, each of which is generated by three generators with relevant structure constants, which may depend on central elements. We construct corresponding Casimir operators and present finite-dimensional unirreps and structure functions via the realizations of these subalgebras in the context of deformed oscillators. By imposing constraints on the structure functions, we obtain the spectrum of the 3D nondegenerate superintegrable system. We also show that this model is multiseparable and admits separation of variables in cylindrical polar and paraboloidal coordinates. We derive the physical spectrum by solving the Schr\"{o}dinger equation of the system and compare the result with those obtained from algebraic derivations.
\end{abstract}

\section{Introduction}
Superintegrable Hamiltonian systems are a very exclusive family of physical systems as they are exactly solvable systems and their symmetries are, in many cases, generated by a nonlinear generalization of Lie algebras \cite{mil13}. In a classical system, a $d$-dimensional dynamical system with Hamiltonian $\mathcal{H}=\frac{1}{2}g^{jk}p_jp_k+V(x)$ and constants of motion $\mathcal{A}_l=f_l(x,p)$, $l=1,\dots,d-1$ is (Liouville) completely integrable if the system allows $d$ integrals including $\mathcal{H}$ that are functionally independent on the phase space, and are in involution $\{\mathcal{H},\mathcal{A}_l\}=0$, $\{\mathcal{A}_l,\mathcal{A}_m\}=0$, $l,m=1,\dots, d-1$. The integrable system is known as superintegrable if it allows additional well-defined constants of motion $\mathcal{B}_m$ on the phase space and they are in involution $\{\mathcal{H},\mathcal{B}_m\}=0$, $m=1,\dots, k$. It is assumed that the set of integrals $\{\mathcal{H},\mathcal{A}_1,\dots,\mathcal{A}_{d-1},\mathcal{B}_1,\dots, \mathcal{B}_k\}$ are functionally independent. The system will be maximally superintegrable if the integrals set has $(2d-1)$ integrals and minimally superintegrable if the set has $d+1$ such integrals. There is a remark that the additional integrals $\mathcal{B}_m$ need not be in involution with $\mathcal{A}_1,\dots, \mathcal{A}_{d-1}$ as well as not with each other.

In quantum mechanics, similar definitions apply with the coordinates $x_i$ and momenta $p_k$, in which they represent as hermitian operators in the Hilbert space satisfying the Heisenberg algebra. Thus, the quantum counterpart of the system is integrable, if there exist $d-1$ well-defined algebraically independent quantum integrals of motion $A_1,\dots, A_{d-1}$ on the Hilbert space that commute with the Hamiltonian operator $H$ and pair-wise with each other, that is, $[H,A_l]=0$ and $[A_l, A_m]=0$ for $1\leq l,m\leq d-1$. The system is superintegrable for the existence of the additional algebraically independent quantum integrals of motion $B_m$ such that $[H, B_m]=0$ for $m=1,\dots, k$. Moreover, the system is known as the quasi-maximally superintegrable system which has $2d-2$ independent constants of motion including the Hamiltonian $H$. The maximally superintegrable system is more special for the existence of a large number of symmetries and for arising many unique properties such as periodic motions and finite closed trajectories or accidental degeneracies of the energy spectrum in classical/(or) quantum mechanics. More exhaustive algebraic descriptions of superintegrable systems in classical and quantum mechanics, symmetry algebras, and their connections to special functions can be found in the review paper \cite{mil13}. Famous examples of superintegrable systems are the Coulomb-Kepler \cite{Foc1, Bar1} and the harmonic oscillator \cite{Jau1, Mos1}.

A systematic algebraic investigation is performed for superintegrable Hamiltonian systems on 2D and 3D Euclidean spaces in \cite{Win1, Mak1, Eva1}. The algebraic computations were more or less completed for the constants of motion which are first- or second-order polynomials of the momenta. Over the year, much work has been done on the complete classifications of second-order classical and quantum superintegrable systems \cite{Ka1, Ka2, Ka3, Ka4, Ka5, Da1}. Nowadays, the search for arbitrary dimensional quantum superintegrable systems and their higher-order constants of motion is a paramount research area (see for examples \cite{Ve1, Ev1, Ro1, Mar1, Mar2, Tr1, Qu1, Ka6, Po1, Ka7}). In the context of the algebraic perspective, the higher-order polynomial algebras with structure constants of certain Casimir invariants are constructed by using the integrals of the $d$-dimensional superintegrable systems \cite{Hoq1, Hoq2, Hoq3, Hoq4, Ch1, Hoq5}. However, the classification of 3D superintegrable Hamiltonian systems is still an active field of research in particular for nondegenerate quantum superintegrable systems and their symmetry algebras \cite{Kal1, das09, das10, es17}. 
The four parameters depending potentials are classified as the nondegenerate potentials, and less than four parameters depending potentials are classified as the degenerate potentials of the 3D superintegrable systems. Such classifications have been explicitly studied in \cite{Kal1, kal03}. The systems with degenerate and nondegenerate potentials were investigated in the seminal paper \cite{Eva1}. It is established that any 3d nondegenerate classical superintegrable system with five second-order constants of motion (including the Hamiltonian) allows an additional integral, which is linearly independent to others. All these integrals of the 3D nondegenerate superintegrable system \cite{Kal1, das09} close to form a parafermionic-like Poisson algebras \cite{Gr1}. The energy spectra of the generalized Coulomb-Kepler system in Euclidean space have been studied using the methods of separation of variables in \cite{Pos1}. The energy eigenvalues of the generalized quantum Kepler-Coulomb system with nondegenerate potentials were calculated algebraically in \cite{das11}. However, the  superintegrable systems with nondegenerate potential, their corresponding quadratic integrals and the symmetry algebras investigation are still interesting problems in quantum mechanics \cite{das09}. We introduce the 3D nondegenerate quantum superintegrable system depending on four parameters, which is known as the {\it{ KKM Potential $V_{IV}$}} \cite{Kal1} with quadratic integrals to present full symmetry quadratic algebra structure. We calculate the energy eigenvalues of the system algebraically. We also show the multiseparability of the system in cylindrical polar and paraboloidal coordinates and solve the Schr\"{o}dinger equation of the system. We compare the result with those obtained from algebraic calculations.

We present this paper in the following form. In section \ref{Se1}, we present a 3D nondegenerate quantum Hamiltonian system in a flat space, and its superintegrability for the set of algebraically independent quadratic integrals. In section \ref{Se2}, we construct the quadratic full symmetry algebra structure generated by the quadratic constants of motion of the system. Section \ref{Se31} contains a brief discussion on the quadratic algebra $Q(3)$ related to the symmetry algebra. In section \ref{Se3}, we recall quadratic subalgebras which are generated by three generators involving structure constants from symmetry algebra and present their corresponding Casimir operators. In section \ref{Se4}, we present the algebraic realizations of the quadratic subalgebras in the context of deformed oscillators of Daskaloyannis's approach \cite{das01, das91} and obtain the energy spectrum of the 3D system. Section \ref{Se7} contains the solutions of the Schr\"{o}dinger equation of the 3D Hamiltonian system in cylindrical polar and paraboloidal coordinates. Section \ref{Se5} contains the concluding remarks.

\section{ The 3D nondegenerate quantum superintegrable system } \label{Se1}
The 3D superintegrable systems with the Hamiltonian 
\begin{eqnarray}
H= \frac{1}{2}\left( p_{x_1}^2+p_{x_2}^2+p_{x_3}^2\right) + V(x_1,x_2, x_3)
\end{eqnarray}
on the flat space have been initially studied in \cite{Eva1}. All these 3D systems with quadratic integrals have been classified in the complex Euclidean space and distinguished to the so-called nondegenerate potentials \cite{Kal1}. These nondegenerate potentials are linear combinations of four parameters, while degenerate potentials depend on less than four parameters. Kalnins, Kress and Miller \cite{Kal1}  also established a general result for the nondegenerate potentials: if $V$ is a 3D nondegenerate potential depending on four parameters considered on a conformally flat space with the metric
\begin{eqnarray}
ds^2=g(x_1,x_2,x_3)(dx_1^2+dx_2^2+dx_3^2),
\end{eqnarray}
then the classical analog of the Hamiltonian associated with the above metric is maximally superintegrable with five functionally independent quadratic integrals $\mathcal{L}=\{S_k : k=1,\dots,5\}$ (including $S_1\equiv H$) and there always exists an extra quadratic integral $S_6$ which is linearly independent to the others. These classical nondegenerate systems and their corresponding quadratic integrals have been performed a parafermionic-like quadratic Poisson algebra on conformally flat space \cite{das09}.
However, the associative quadratic ternary symmetry algebras for the nondegenerate superintegrable systems in quantum mechanics and their degeneracy of the energy spectra remain largely unknown. Their analytic solutions to the Schr\"{o}dinger equations via separation of variables would be of much interest. At the first attempt, we thus introduce a nondegenerate quantum system, which is known as the {\it KKM potential $V_{IV}$}, with the Hamiltonian operator in real Euclidean space $\mathbb{E}^3$ \cite{Kal1}, 
\begin{eqnarray}
H= p_{x_1}^2+p_{x_2}^2+p_{x_3}^2+ c_1 (4 x_1^2+x_2^2+x_3^2) + c_2  x_1+ \frac{c_3}{x_2^2}+\frac{c_4}{x_3^2},\label{Ham1}
\end{eqnarray}
where $p_{x_i}=-i\frac{\partial}{\partial x_i}$ and we set $m=\hbar=1$. It is remarked that $c_2$ can be eliminated by a shift of coordinates under the condition that $c_1$ is not zero and it is possible to reduce the potential to 3 parameters, in this case, the algebra must transform correspondingly. In this paper, we present the full symmetry quadratic algebra of the 3D nondegenerate quantum superintegrable system (\ref{Ham1}) and obtain the energy spectrum applying the Daskaloyannis deformed oscillator algebra approach \cite{das01} on the symmetry algebras.

The quantum Hamiltonian system (\ref{Ham1}) has the following four algebraically independent quadratic integrals,
\begin{eqnarray}
 &&A_1= p_{x_1}^2+ 4 c_1 x_1^2+ c_2 x_1, \quad
A_2=p_{x_2}^2 + c_1 x_2^2+\frac {c_3}{x_2^2}, 
\nonumber\\&&
B_1= J_2 p_{x_3}+p_{x_3} J_2+ 2c_1 {x_1 x_3^2}+\frac{c_2 x_3^2}{2}-\frac{2c_4 x_1}{x_3^2},
\quad
 B_2= J_1^2 +\frac{c_3 x_3^2}{x_2^2}+\frac{c_4 x_2^2}{x_3^2},
\end{eqnarray} 
and one additional quadratic integral,
\begin{eqnarray}
F= p_{x_2} J_3 +J_3 p_{x_2}-2 c_1 x_1 x_2^2-\frac{c_2 x_2^2}{2}+\frac{2 c_3 x_1}{x_2^2},
\end{eqnarray}
where 
\begin{eqnarray}
J_{1}=x_2  p_{x_3}- x_3 p_{x_2},  \quad J_{2}=x_3  p_{x_1}- x_1 p_{x_3}, \quad  J_{3}= x_1  p_{x_2}- x_2 p_{x_1}.
\end{eqnarray}
All these integrals are linearly independent including the Hamiltonian $H$. The Hamiltonian system is maximally superintegrable. It can be proved by the following commutation relations
\begin{eqnarray}
&&  [A_i,H]=0,\quad [B_i,H]=0,\quad i=1,2,\quad [H,F]=0.
\end{eqnarray}
We also found that
\begin{eqnarray}
[A_1,B_2]=0,\quad [ A_1,A_2]=0, \quad  [A_2,B_1]=0.
\end{eqnarray}
The above commutativity relations can be expressed as the following diagram to easily understand, 

\begin{eqnarray}
\begin{xy}
(0,0)*+{B_{2}}="m"; (0,40)*+{A_1}="q"; (40,40)*+{A_2}="p"; (40,0)*+{B_{1}}="n"; (20,20)*+{H}="r"; (20,0)*+{F}="l"; 
"m";"q"**\dir{--};
"q";"p"**\dir{--};
"p";"n"**\dir{--};
"m";"r"**\dir{--};
"q";"r"**\dir{--};
"p";"r"**\dir{--};
"n";"r"**\dir{--};
"l";"r"**\dir{--};
\end{xy}
\end{eqnarray}
where the dashed lines indicate that the commutator of the corresponding integrals is zero and the absence of dashed lines among the integrals means the commutator is nonzero. The presence of $F$ ensures that the integrals generate a ternary type quadratic algebra involving six generators including the Hamiltonian operator.  We can also define
\begin{eqnarray}
C_1 = [A_1,B_1],  \quad C_2= [A_2,B_2], \quad D=[B_1,B_2]. \label{E1.8}
\end{eqnarray}
It is shown that the new integrals of motion $C_1$ and $C_2$ are cubic functions of momenta, which can not be expressed as a polynomial function in terms of other integrals of motion that are the second structure of momenta. We may also define
\begin{eqnarray}
E_1=[A_1,F],\quad E_2=[A_2,F],\quad E_3=[B_1,F],\quad E_4=[B_2,F].
\end{eqnarray} 

\section{\textbf{Quadratic symmetry algebra}} \label{Se2}
We now derive the full quadratic symmetry algebra of the quantum superintegrable Hamiltonian system (\ref{Ham1}). After a long direct computation and using different commutation relations and Jacobi identities, the six integrals of motion including the Hamiltonian $H$ close to form the following quadratic symmetry algebra,
\begin{eqnarray}
&& [ A_1,C_1] = 4c_2A_1+16 c_1 B_1+4 c_2(A_2-H),\label{SA1}
\\&&
[ B_1,C_1] =  24 A_1^2 + 32 (A_2-H)A_1 - 4 c_2B_1+ 8 H^2-16 H A_2 \nonumber\\&&\qquad \qquad\quad -8c_1( 4 c_4 -3)+8 A_2^2,\label{SA2}
\\&&
[ A_2,C_2] =  8 A_2^2+8(A_1-H)A_2 +8 c_1 (2B_2+1),\label{SA3}
\\&&
[ B_2,C_2] = -16(c_3+c_4-1) A_2-8\{A_2, B_2\} -8(A_1 
 -H)B_2 \nonumber\\&&\qquad \qquad\qquad -8(2 c_3-1) A_1 +8(2 c_3-1)H,\label{SA4}
\\&&
[A_1,D]=8A_2 B_1-8FA_1-8A_2F+8HF,
\\&&
[C_1,F]=8H A_2-8 A_2 A_1- 8 A_2^2-16 c_1 B_2 - 8 c_1,
\\
&&
[C_1,B_2]=8 A_2 B_1-8 A_1 F- 8 A_2 F + 8 HF,
\\&&
[E_1,A_2]=16 c_1F+4c_2 A_2,
\\&&
[E_1,B_2]=  8H F -8 F A_1-8A_2 F +8 B_1 A_2,
\\&&
[E_1,F]= 16 A_2 A_1- 4 c_2 F - 8 A_2^2 + 8c_1(4 c_3-3),
\\&&
[E_1, A_1]= -16 c_1 F- 4 c_2 A_2,
\\
&& 
[C_2,B_1]=4 c_2 B_2-8 F A_2-8 A_1 F + 8 HF +2 c_2,
\\&&
[E_2,A_2]= -4 c_2 A_2-16 c_1 F,
\\&&
[E_2,B_1]=8 A_2^2+8(A_1-H)A_2 +16 c_1 B_2+8 c_1,
\\&&
[E_2,B_2]= 8 A_2 F+ 8 A_1 F- 8 HF - 8 B_1 A_2,
\\&&
[E_2,F]=  8 A_2^2-16 A_1 A_2+4 c_2F -8c_1(4 c_3-3),
\end{eqnarray}
\begin{eqnarray}
&&
[B_1,D]= 8 FB_1-8 ( A_2+3 A_1-H) B_2-8(2c_4-1)A_2 - 12 A_1+4 H,
\\&&
[D,B_2]= 8(B_1 B_2+B_2 B_1)+8 F B_2+8(2 c_3-1)B_1 +8(2 c_4-1)F-8 B_1 B_2,
\\&&
[E_3,B_1]= 8(A_1+A_2-H)F-4c_2 B_2- 2 c_2,
\\
&&
[E_3,B_2]= 8FB_1-8(2c_3-1)A_1-16(c_3+c_4-1)A_2
  +8(2c_3-1)H \nonumber\\&&\qquad \qquad\qquad -8A_1B_2-16A_2B_2+8B_2-8B_1F, 
\\&&
[E_3,F]=8 A_2 B_1-4c_2 B_2 - 2 c_2,
\\&&
[E_4,A_1]= -8(A_1+A_2-H)F+ 8 B_1 A_2,
\\&&
[E_4,B_2]= -8(B_2 F+F B_2)+8 F B_2-8 (2 c_4-1)F- 8 B_1 B_2 -8 (2 c_3-1) B_1,
\\&&
[E_4,F]=-8(2 A_1-A_2)B_2+ 8 B_1 F- 4 (4 c_3-3)H +4(4 c_3-5)A_1+8(2 c_3-1)A_2.
\end{eqnarray}
We can also present a second algebra in terms of $C_1$, $C_2$ and $D$ with coefficients in linear combinations of integrals $A_1, A_2, B_1, B_2, F, H$ as
\begin{eqnarray}
&& [C_1,C_2]=8A_2C_1-4c_2C_2-16c_1D,\nonumber\\&&
[C_1,D]=8FC_1-8A_2C_2-24A_1C_2+8HC_2+4c_2D,\\&&
[C_2,D]=-8B_2C_1-8FC_2+8A_2D-8(2c_3-1)C_1.\nonumber
\end{eqnarray}
To the observation, the relations ((\ref{SA1}) - (\ref{SA2})) and ((\ref{SA3})- (\ref{SA4})), respectively, involving the integrals set $\{A_1, B_1, C_1\}$ and $\{A_2, B_2, C_2\}$, defined by the subalgebras, $Q_1(3)$ and $Q_2(3)$, have a connection to the quadratic algebra $Q(3)$ of 2D superintegrable systems with quadratic integrals of motion \cite{das01}.
It is stimulating to see the subalgebras $Q_i(3), i=1,2$ are embedded in the symmetry algebra of the 3D nondegenerate superintegrable system (\ref{Ham1}).
  In the following section, we consider the subalgebras $Q_i(3), i=1,2$ to calculate the spectrum of the system (\ref{Ham1}) by using the Daskaloyannis approach and the deformed oscillator algebra realizations \cite{das01,das91}.

\section{The quadratic algebra $Q(3)$}\label{Se31}
In the above section, we successfully obtain the symmetry algebra structures of the 3D superintegrable system (\ref{Ham1}) for nondegenerate potential. We now have to calculate the energy spectrum of the system based on the symmetry algebra. In order to derive the spectrum algebraically, we demonstrate the existence of a set of subalgebra structures $Q_i(3), i=1,2$, involving three generators and compared them with the quadratic algebra $Q(3)$ presented by Daskaloyannis in the context of 2D superintegrable systems \cite{das01}. We recall briefly this algebraic method for two subalgebras $Q_i(3), i=1,2$ which involves three operators $\{\mathcal{A}_i, \mathcal{B}_i, \mathcal{C}_i\}$ for $i=1,2$ and $[\mathcal{A}_i,\mathcal{A}_j]=0$, for all $i,j$ \cite{Hoq5}. They are close to form the following quadratic algebras,
\begin{eqnarray}
[\mathcal{A}_i,\mathcal{B}_i]&=&\mathcal{C}_i,\nonumber\\
{}[\mathcal{A}_i,\mathcal{C}_i]&=&\alpha_i \mathcal{A}_i^2+\gamma_i \{\mathcal{A}_i,\mathcal{B}_i\}+\delta_i \mathcal{A}_i +\epsilon_i \mathcal{B}_i +\zeta_i,\nonumber\\
{}[\mathcal{B}_i,\mathcal{C}_i]&=&a_i \mathcal{A}_i^2-\gamma_i \mathcal{B}_i^2-\alpha_i \{\mathcal{A}_i,\mathcal{B}_i\} +d_i \mathcal{A}_i-\delta_i \mathcal{B}_i+z_i,    \label{E19}
\end{eqnarray}
where $i=1,2$. The coefficients $\alpha_i$, $\gamma_i$, $a_i$ are constants and $d_i$, $\delta_i$, $\epsilon_i$, $\zeta_i$,  $z_i$ are polynomials of central elements: the Hamiltonian $H$ and the generator $\mathcal{A}_j$ of the $j$-th subalgebra, which commutes with the generators of the $i$-th subalgebra. 
The generators $\{\mathcal{A}_i,\mathcal{B}_i,\mathcal{C}_i\}$ of the subalgebras $Q_i(3), i=1,2$ form a Casimir invariant
\begin{align}
\mathcal{K}_i&=\mathcal{C}_i^{2}-\alpha_i\{\mathcal{A}_i^2,\mathcal{B}_i\}-\gamma_i \{\mathcal{A}_i,\mathcal{B}_i^{2}\}+(\alpha_i \gamma_i-\delta_i)\{\mathcal{A}_i,\mathcal{B}_i\}+ (\gamma_i^2-\epsilon_i) \mathcal{B}_i^{2} \nonumber\\
&+ (\gamma_i\delta_i-2\zeta_i) \mathcal{B}_i+\frac{2a_i}{3}\mathcal{A}_i^3 +\left(d_i+\frac{a_i\gamma_i}{3}+\alpha_i^2\right) \mathcal{A}_i^{2} +\left(\frac{a_i\epsilon_i}{3}+\alpha_i\delta_i+2z_i\right) \mathcal{A}_i . \label{E20}
\end{align}
It is pointed out that this Casimir invariant is possible to reform in terms of only central elements of the corresponding subalgebras. To determine the spectrum of the Hamiltonian operator $H$, the algebra $ Q_i(3)$ (\ref{E19}) $i=1,2$ needs to realize in terms of the deformed oscillator algebras \cite{das01, das91},
\begin{equation}
[\aleph_i,b_i^{\dagger}]=b_i^{\dagger},\quad [\aleph_i,b_i]=-b_i,\quad b_ib_i^{\dagger}=\phi (\aleph_i+1),\quad b_i^{\dagger} b_i=\phi(\aleph_i),  \label{Deform}
\end{equation}
and the structure function $\phi$ for $\gamma_i \neq 0$ is given by
\begin{align}\nonumber
 \phi_i(n_i) &=\gamma_i^8 (3 \alpha_i^2 {+ }4 a_i \gamma_i )[2 (n_i{+}u_i){-}3]^2 [2 (n_i{+}u_i){-}1]^4 [2 (n_i{+}u_i){+}1]^2-3072 \gamma_i^6  \mathcal{K}_i[2 (n_i{+}u_i){-}1]^2    \\ \nonumber
&-48  \gamma_i^6(\alpha_i^2 \epsilon_i{-}\alpha_i  \gamma_i  \delta_i{+} a_i \gamma_i  \epsilon_i{-}\gamma_i^2 d_i)[2 (n_i{+}u_i){-}1]^4[2 (n_i{+}u_i){+}1]^2[2 (n_i{+}u_i){-}3] \\
&+32 \gamma_i^4  \left(3 \alpha_i^2 \epsilon_i^2{+}4 \alpha_i  \gamma_i^2 \zeta_i {-}6 \alpha_i  \gamma_i  \delta_i  \epsilon_i {+}2 a_i \gamma_i  \epsilon_i^2{+}2 \gamma_i^2 \delta_i^2{-}4 \gamma_i^2 d_i \epsilon_i{+}8 \gamma_i^3 z_i\right) \times \\ \nonumber
&[2 (n_i{+}u_i){-}1]^2[12 (n_i{+}u_i)^2{-}12 (n_i{+}u_i){-}1]+ 768 (\alpha_i  \epsilon_i^2+4 \gamma_i^2 \zeta_i -2 \gamma_i  \delta_i  \epsilon_i )^2 \\ \nonumber
&-256\gamma_i^2[2 (n_i{+}u_i){-}1]^2(3 \alpha_i^2 \epsilon_i^3{+}4 \alpha_i  \gamma_i^4 \zeta_i {+}12 \alpha_i  \gamma_i^2 \zeta_i  \epsilon_i{-}9 \alpha_i  \gamma_i  \delta_i  \epsilon_i^2{+}a_i \gamma_i  \epsilon_i^3{+}2 \gamma_i^4 \delta_i^2 \\ \nonumber
&-12 \gamma_i^3 \delta_i  \zeta_i{+}6 \gamma_i^2 \delta_i^2 \epsilon_i{+}2 \gamma_i^4 d_i \epsilon_i{-}3 \gamma_i^2 d_i \epsilon_i^2{-}4 \gamma_i^5 z_i{+}12 \gamma_i^3 z_i \epsilon_i )
\end{align}
and the eigenvalues of the operator $\mathcal{A}_i$, 
\begin{eqnarray}
 e(\mathcal{A}_i)=\mathcal{A}_i(q_i)=\sqrt{\epsilon_i}(q_i+u_i), \qquad \gamma_i=0,\quad \epsilon_i \neq 0;\label{E3.22}
\end{eqnarray} 
 and for the case $\gamma_i=0$, $\epsilon_i \neq 0$, the structure function is given by
\begin{align}
 \Phi_i(n_i)  &=  \frac{1}{4}\left[{-}\frac{\mathcal{K}_i}{\epsilon_i}{-}\frac{z_i}{\sqrt{\epsilon_i}}{-}\frac{\delta_i}{\sqrt{\epsilon_i}}\frac{\zeta_i}{\epsilon_i}{+}\left(\frac{\zeta_i}{\epsilon_i}\right)^2\right]  \nonumber\\
 & -\frac{1}{12}\left[3d_i{-}a_i\sqrt{\epsilon_i}{-}3\alpha_i \frac{\delta_i}{\sqrt{\epsilon_i}}{+}3 \frac{\delta_i^2}{\epsilon_i}{-}6\frac{z_i}{\sqrt{\epsilon_i}}{+}6\alpha_i\frac{\zeta_i}{\epsilon_i}-6\frac{\delta_i}{\sqrt{\epsilon_i}}\frac{\zeta_i}{\epsilon_i}\right](n_i{+}u_i)\nonumber\\
&+\frac{1}{4}\left[ \alpha_i^2{+}d_i{-}a_i\sqrt{\epsilon_i}{-}3\alpha_i\frac{\delta_i}{\sqrt{\epsilon_i}}{+}\frac{\delta_i^2}{\epsilon_i}{+}2\alpha_i\frac{\zeta_i}{\epsilon_i}  \right](n_i{+}u_i)^2
 \nonumber\\
 &-\frac{1}{6}\left[3\alpha_i^2{-}a_i\sqrt{\epsilon_i}{-}3\alpha_i\frac{\delta_i}{\sqrt{\epsilon_i}}  \right](n_i{+}u_i)^3{+}\frac{1}{4}\alpha^2(n_i{+}u_i)^4
\end{align}
 and the eigenvalues of the operator $\mathcal{A}_i$, 
\begin{eqnarray}
e(\mathcal{A}_i)=\mathcal{A}_i(q_i)=\frac{\gamma_i}{2}\left((q_i+u_i)^2-\frac{\epsilon_i}{\gamma_i^2}-\frac{1}{4}\right), \quad \gamma_i \neq 0.\label{E3.21}
\end{eqnarray}

\section{The subalgebras $Q_i(3)$, $i=1,2$} \label{Se3}
The relations ((\ref{SA1}) - (\ref{SA2})) and ((\ref{SA3})- (\ref{SA4})) of the quadratic symmetry algebras formed similar quadratic structure $Q(3)$ (\ref{E19}) involving three generators sets $\{A_1, B_1, C_1\}$ and $\{A_2, B_2, C_2\}$.
The subalgebras $Q_i(3), i=1,2$ can be presented in the following diagrams, 
\begin{eqnarray}
\begin{xy}
(0,0)*+{A_1}="m"; (20,0)*+{A_2}="q";  (40,0)*+{B_1}="n"; (20,20)*+{H}="r"; 
"m";"q"**\dir{--}; 
"q";"n"**\dir{--}; 
"n";"r"**\dir{--};
"r";"m"**\dir{--};
"r";"q"**\dir{--};
\end{xy} 
\qquad\quad
\begin{xy}
(0,0)*+{A_2}="m"; (20,0)*+{A_1}="q";  (40,0)*+{B_2}="n"; (20,20)*+{H}="r"; 
"m";"q"**\dir{--}; 
"q";"n"**\dir{--}; 
"n";"r"**\dir{--};
"r";"m"**\dir{--};
"r";"q"**\dir{--};
\end{xy}
\end{eqnarray}
The left figure shows that $A_2$ and $H$ are central elements and the right figure shows that $A_1$ and $H$ are central elements of the corresponding subalgebra structures. It is seen as a fact that one integral plays a role as a generator in a subalgebra structure while it plays a role as a central element in another subalgebra structure. In account to obtain the spectrum, we manipulate the subalgebras $Q_i(3)$, $i=1,2$ and it is clear that each of these subalgebras has a relationship with the quadratic algebra (\ref{E19}) and Casimir operator (\ref{E20}) presented in  \cite{das01} for the 2D superintegrable system. We rewrite the relations ((\ref{SA1}) - (\ref{SA2})) as the subalgebra structure $Q_1(3)$,
\begin{eqnarray}
&&  [A_1, B_1]= C_1, \nonumber \\&&
[A_1, C_1]=4c_2A_1+16c_1B_1+4c_2(A_2-H),  \label{Q1}        \\&&
[B_1, C_1]=24A_1^2+32(A_2-H)A_1-4c_2B_1+8H^2 -16HA_2 \nonumber\\&&  \qquad  \qquad -8c_1(4c_4-3)+8A_2^2,\nonumber
\end{eqnarray}
and the relations ((\ref{SA3})- (\ref{SA4})) as the subalgebra structure $Q_2(3)$,
\begin{eqnarray}
&&  [A_2, B_2]= C_2, \nonumber \\&&
[A_2, C_2]=8A_2^2+8(A_1-H)A_2+16c_1B_2+8c_1,\label{Q2} \\&&
[B_2, C_2]= -8\{A_2,B_2\}-16(c_3+c_4-1)A_2 -8(A_1-H)B_2\nonumber\\&&\qquad\qquad +8(2c_3-1)(H-A_1).\nonumber
\end{eqnarray}
There are Casimir operators $K_1$ of $Q_1(3)$ and $K_2$ of $Q_2(3)$ satisfying $[K_1, A_1]=0=[K_1,B_1]$ and $[K_2, A_2]=0=[K_2,B_2]$, respectively,
\begin{eqnarray}
&& K_1=C_1^2-4c_2\{A_1,B_1\}-16c_1B_1^2-8c_2(A_2-H)B_1+16A_1^3\nonumber\\&&  \qquad +32(A_2-H)A_1^2  +[128c_1+16H^2-32HA_2 -16c_1(4c_4-3) +16A_2^2]A_1,\label{K1}
\end{eqnarray}
and 
\begin{eqnarray}
&&  K_2=C_2^2-8\{A_2^2,B_2\}-8(A_1-H)\{A_2,B_2\}-16 c_1 B_2^2  -16 c_1 B_2 \nonumber\\&&\qquad -16(c_3+c_4-5) A_2^2-16(2c_2-5)(A_1-H)A_2.\label{K2}
\end{eqnarray}
Moreover, the quadratic subalgebras $Q_1(3)$ and $Q_2(3)$ possess corresponding Casimir invariants in terms of only central elements in the following forms, respectively,
\begin{eqnarray}
K_1'= 128c_1H - 128c_1A_2-3c_2^2+4c_2^2c_4,\label{K1'}
\end{eqnarray}
and
\begin{eqnarray}
&& K_2'=4(4c_3-3)(H-A_1)^2  -16 (2 c_1 - 3 c_1 c_3 -3 c_1 c_4 + 4 c_1 c_3 c_4).\label{K2'}
\end{eqnarray} 
It is seen that the Casimir operator $K'_1$ depends on only central elements $H$ and $A_2$ for the subalgebra $Q_1(3)$, and the Casimir operator $K'_2$ depends on only central elements $H$ and $A_1$ of the subalgebra $Q_2(3)$. These two forms of the Casimir invariants will be used to realize these subalgebras in terms of the deformed oscillator algebras (\ref{Deform}).

\section{Deformed oscillators realizations and energy spectrum} \label{Se4}
In order to obtain the energy spectrum of the superintegrable system (\ref{Ham1}), we realize the subalgebra structures $Q_1(3)$ and  $Q_2(3)$ in terms of deformed oscillator algebra \cite{das01, das91} $\{\aleph_i, b^{\dagger}_i,b_i\}$ of (\ref{Deform}) with satisfying the real-valued function,
\begin{eqnarray}
\phi(0)=0, \quad \phi(n_i)>0, \quad \forall n_i >0.
\end{eqnarray} 
The $\phi(n_i)$ is known as structure function. We first investigate the realization of the quadratic subalgebra structure $Q_1(3)$ (\ref{Q1}).
The realization of $Q_1(3)$ is of the form $A_1=A_1(\aleph_1)$, $B_1=b_1(\aleph_1)+b_1^{\dagger}\rho_1(\aleph_1)+\rho_1(\aleph_1)b_1$, where $A_1(x)$, $b_1(x)$ and $\rho_1(x)$ are functions to lead the following forms \cite{das01},
\begin{eqnarray}
&& A_1(\aleph_1)=\sqrt{16c_1}(\aleph_1+u_1),\label{EA1} \\&&
b_1(\aleph_1)= -\frac{c_2}{\sqrt{c_1}}(\aleph_1+u_1)-\frac{c_2(A_2-H)}{4c_1},\\&&
\rho_1(\aleph_1)=1,
\end{eqnarray}
where $u_1$ is an arbitrary constant to be determined later. The following structure function $\phi(n_1,u_1, H)$ of the subalgebra (\ref{Q1}) is constructed by using the deformed oscillators (\ref{Deform}) and the Casimir invariants (\ref{K1}) and (\ref{K1'}),
\begin{eqnarray}
&&\phi_1(n_1, u_1, H)=\frac{1}{1024m_1^5}\bigg[\left(A_2-H\right)-m_1  \left(2+m_4-4\left(n_1+u_1 \right)\right)\bigg]\nonumber\\&&\qquad\qquad\qquad\qquad \bigg[\left(A_2-H\right)+m_1 \left(-2+m_4+4\left(n_1+u_1 \right)\right)\bigg] \nonumber\\&&\qquad\qquad\qquad\qquad \bigg[m_2^2+32m_1^3\left(-1+2\left(n_1+u_1\right)\right)\bigg],
\label{ST1}
\end{eqnarray}
where $m_1^2=c_1$, $m_2=c_2$, $m_4^2=4c_4+1$.
To obtain the eigenvalues of the central element $A_2$ of this subalgebra and the values of parameter $u_1$ by requiring that the unitary representations (unirreps) to be a finite, we should impose the following constraints on the structure function:
\begin{eqnarray}
\phi(p_1+1; u_1, E)=0; \quad \phi(0; u_1, E)=0; \quad \phi(x)>0, \quad \forall x>0,\label{Cns}
\end{eqnarray}
where $p_1$ is a positive integer. We also replace the eigenvalue $E$ of $H$ in $\phi_1(n_1, u_1, H)$. The constraints guarantee the structure functions are finite $(p_1+1)$-dimensional unirreps. We solve the constraints (\ref{Cns}), which give information of the eigenvalues $e(A_2)$ of $A_2$ and the values of the constant $u_1$. Imposing the condition (\ref{Cns}) to the structure functions (\ref{ST1}) for unirreps of finite-dimensional $(p_1+1)$ and positive values of the structure function, we obtain the solutions with $\varepsilon_1= +1$, $\varepsilon_2=\pm 1$, $\varepsilon_3=\pm 1$,
\begin{eqnarray}
&& u_{1}=\frac{1}{2}-\frac{ m_2^2}{64 m_1^3},\quad \text{or}\quad u_{1}=\frac{1}{4 m_1}\left[E +2 m_1+\varepsilon_1 m_1m_4- e(A_2) \right],  \label{U1}\\&&
 e(A_2)= 4\varepsilon_2 m_1(p_1+1)+E+\varepsilon_3 m_1 m_4+\frac{m_2^2}{16m_1^2}. \label{A2}
\end{eqnarray}
It is remarked that the subalgebra $Q_1(3)$ is the algebra of the superintegrable two-dimensional subsystem, depending on the variables $x_1, x_3$. The above computation shows that the energy of the subsystem equal to $E-e(A_2)$ and the quadratic algebra depends on $H-A_2$, it is thus clearly understandable the existence of the Casimir invariant (\ref{K1'}). A similar reason is applicable for the subalgebra $Q_2(3)$ and the Casimir invariant (\ref{K2'}). We now obtain  the eigenvalues of $A_1$ from the relations (\ref{EA1}) and (\ref{U1}) as follows,
\begin{eqnarray}
e(A_1)=2m_1(2n_1 +1)-\frac{ m_2^2}{16 m_1^2}, \quad \text{or}\quad e(A_1)=2m_1(2n_1 +1)+E +\varepsilon_1 m_1m_4 - e(A_2).\label{A1}
\end{eqnarray} 
We now turn to the quadratic subalgebra structure $Q_2(3)$ (\ref{Q2}). Similar to $Q_1(3)$, the realizations of $Q_2(3)$ present the functions
\begin{eqnarray}
&& A_2(\aleph_2)=\sqrt{16c_1}(\aleph_2+u_2),\label{EA2} \\&&
b_2(\aleph_2)= -8(\aleph_2+u_2)^2-\frac{2(A_1-H)}{\sqrt{c_1}}(\aleph_2+u_2)-\frac{1}{2},\\&&
\rho_2(\aleph_2)=1
\end{eqnarray}
and the structure function
\begin{eqnarray}
&&\phi(n_2, u_2,H)=\frac{1}{256m_1^4}[-2-m_3+4(n_2+u_2)][-2+m_3+4(n_2+u_2)]\nonumber\\&&\qquad\qquad\qquad [A_1m_1-m_1H-m_1^2m_4+m_1^2(-2+4(n_2+u_2)]\nonumber\\&&\qquad\qquad\qquad [A_1m_1-m_1H+m_1^2m_4+m_1^2(-2+4(n_2+u_2)],\label{ST2}
\end{eqnarray}
where  $m_3^2=4c_3+1$.
We now impose the constraints (\ref{Cns}) on the structure function (\ref{ST2}) for unirreps of finite-dimensional $(p_2+1)$ and positive values of the structure function, giving the following solutions, 
\begin{eqnarray}
&& u_2=\frac{1}{2}+\frac{\varepsilon_1 m_3}{4},\quad \text{or}\quad u_{2}= \frac{1}{4m_1}\left[E + 2 m_1  +\varepsilon_1 m_1 m_4 -e(A_1)\right], \label{U21} \\&&
 e(A_1)=4 \varepsilon_1 m_1 (p_2 + 1)+ E+\varepsilon_2 m_1 m_3+\varepsilon_3 m_1 m_4,\label{SA5}
\end{eqnarray}
where $\varepsilon_1= +1$, $\varepsilon_2=\pm 1$ and $\varepsilon_3=\pm 1$.
Similar, from the relations (\ref{EA2}) and (\ref{U21}), we obtain the eigenvalues $e(A_2)$ of $A_2$ as
\begin{eqnarray}
e(A_2)=2m_1(2n_2+1)+\varepsilon_1 m_1 m_3,  \quad e(A_2)=2m_1(2n_2+1) + E +\varepsilon_1 m_1 m_4 - e(A_1).\label{SA6}
\end{eqnarray}
The energy eigenvalues of the superintegrable system (\ref{Ham1}) are calculated using the  relations (\ref{A1}),  (\ref{SA5}), and (\ref{A2}),  (\ref{SA6}), and choosing the suitable sign of $\varepsilon_i, i=1,2,3$ for positive energy levels,
\begin{eqnarray}
&& E = 4 (p_2+1)m_1+2(2n_1+1)m_1+ m_1m_3+ m_1m_4-\frac{m_2^2}{16m_1^2},\label{E1}
  \\&&
E = 4 (p_1+1)m_1+2(2n_2+1)m_1+ m_1m_3+ m_1m_4-\frac{m_2^2}{16m_1^2}.\label{E2}
\end{eqnarray}
It is a fact that the elimination of the energy $E$ from the above relations (\ref{E1}) and (\ref{E2}) leads to a relation,
\begin{eqnarray}
p_1-p_2=n_1-n_2,
\end{eqnarray}
which is valid, because the two deformed oscillators are treated as independent ones,
\begin{eqnarray}
n_1=0,1,2,\dots,p_1,\quad \text{and}\quad n_2=0,1,2,\dots,p_2.
\end{eqnarray}
The mean value of the relations (\ref{E1}) and (\ref{E2}) reduce to the energy eigenvalues of the superintegrable Hamiltonian system (\ref{Ham1}),
\begin{eqnarray}
 E = 2 (p_1+p_2+2)m_1+2(n_1+n_2+1)m_1+ m_1m_3+ m_1m_4-\frac{m_2^2}{16m_1^2}.\label{Eal1}
\end{eqnarray}
It is a fact that the quadratic subalgebra structures of the symmetry algebra provide us the energy spectrum for the 3D nondegenerate potential of  the maximally quantum superintegrable system (\ref{Ham1}) purely algebraic computations. It is shown that the energy spectrum of the system in the algebraic investigation depends only on differential operators and their corresponding operator algebra in symmetry forms without knowledge of wave functions and calculus.

\section{Separation of variables}\label{Se7}

We now demonstrate analytic calculations of the 3D superintegrable system (\ref{Ham1}) via separation of variables in the cylindrical polar and paraboloidal coordinates. The results will be compared with those obtained from algebraic derivations.

\subsection{Cylindrical polar coordinates}
The cylindrical polar coordinates are given by
\begin{eqnarray}
x_1=z,\quad x_2=\rho \sin^2{\theta},\quad x_3=\rho \cos^2{\theta},
\end{eqnarray}
where $\rho >0$, $-\infty<z<\infty$ and $\theta\in [0,2\pi]$ \cite{Pos1}. The Schrodinger equation $H\psi =
E\psi$ of the system (\ref{Ham1}) in the coordinates can be expressed as
\begin{eqnarray}
&&\bigg[-\left(\frac{\partial^2}{\partial z^2}+\frac{\partial^2}{\partial \rho^2}+\frac{1}{\rho} \frac{\partial}{\partial \rho}+\frac{1}{\rho^2} \frac{\partial^2}{\partial \theta^2}\right)+c_1(4z^2+\rho^2)+c_2 z+\frac{c_3}{\rho^2 \sin^2\theta} \nonumber\\&&\qquad\qquad +\frac{c_4}{\rho^2 \cos^2\theta}-E\bigg]\psi(\rho,z,\theta)=0. \label{EAA}
\end{eqnarray}
The separation of variables of (\ref{EAA}) 
\begin{eqnarray}
\psi(\rho,z,\theta)=R(\rho,z) Y(\theta)
\end{eqnarray}
gives rise to the angular and radial parts with separation constant $A$,
\begin{eqnarray}
\bigg[- \frac{\partial^2}{\partial \theta^2}+\frac{c_3}{ \sin^2\theta}+\frac{c_4}{\cos^2\theta}-A\bigg]Y(\theta)=0,\label{EAAAA}
\end{eqnarray}
\begin{eqnarray}
\bigg[-\left(\frac{\partial^2}{\partial z^2}+\frac{\partial^2}{\partial \rho^2}+\frac{1}{\rho} \frac{\partial}{\partial \rho}\right)+c_1(4z^2+\rho^2)+c_2 z-E+\frac{A}{\rho^2}\bigg]R(\rho,z)=0.\label{EAAA}
\end{eqnarray}

We now take the equation (\ref{EAAA}) to separate the variables
\begin{eqnarray}
R(\rho,z)=G(\rho) F(z)
\end{eqnarray}
we obtain,
\begin{eqnarray}
\bigg[-\frac{\partial^2}{\partial z^2}+c_1 4z^2+c_2 z-E-A_1\bigg]F(z)=0,\label{EABBBB}
\end{eqnarray}
\begin{eqnarray}
\bigg[-\frac{\partial^2}{\partial \rho^2}-\frac{1}{\rho} \frac{\partial}{\partial \rho}+c_1\rho^2+\frac{A}{\rho^2}+A_1\bigg]G(\rho)=0,\label{EABB}
\end{eqnarray}
where $A_1$ is a separation constant.

We now turn to (\ref{EAAAA}), which converts to, by setting $\upsilon=\sin^2{\theta}$ and $g(\upsilon)=\upsilon^{\frac{1}{4}(1+2\gamma_3)} (1-\upsilon)^{\frac{1}{4}(1+2\gamma_4)}{g_1}(\upsilon) $,
\begin{eqnarray}
\upsilon(1-\upsilon) {g_1}^{''}(\upsilon)+\bigg[\left(1\pm {\gamma}_3\right)-\left(1+1\pm \gamma_3\pm \gamma_4\right)\upsilon\bigg]{g_1}^{'}(\upsilon)-\nonumber\\ \bigg[\left(\frac{1}{4}\pm \frac{\gamma_3}{2}+\frac{1}{4}\pm \frac{\gamma_4}{2}\right)^2- \frac{A}{4} \bigg]g_1(\upsilon)=0,\label{EAAAAA}
\end{eqnarray}
where $ \gamma_3=\pm \frac{1}{2} \sqrt{1+4 c_3}$ and  $ \gamma_4=\pm \frac{1}{2} \sqrt{1+4 c_4}.$ By comparing with the Jacobi differential equation \cite{magn1} 
\begin{eqnarray}
z\left(1-z\right)y^{''}+\left[\gamma-\left(\alpha+1\right)z\right]y^{'}+n\left(\alpha+n\right)y=0,\label{EAB}
\end{eqnarray}
 we find the separation constant 
\begin{eqnarray}
A=\left(2n\pm \gamma_3\pm \gamma_4+1\right)^2,
\end{eqnarray}
where $n$ is positive integers.
Hence we have the solutions of (\ref{EAAAAA}) as follows 
\begin{eqnarray}
 Y\left(\theta\right)&&=({\sin^2{\theta}})^{\frac{1}{4}\pm \frac{\gamma_3}{2}} (1-\sin^2{\theta})^{\frac{1}{4}\pm \frac{\gamma_4}{2}}\nonumber\\&& \times \bigg[C_{12} F_1\left(-n, n+1\pm \gamma_3\pm \gamma_4,1\pm \gamma_3,\sin^2{\theta}\right) -\left(-1\right)^{-(1\pm \gamma_3)} \left(\sin^2 \theta\right)^{\pm {\gamma_3}}  \nonumber\\&&\times C_{22} F_1\bigg[(\pm \gamma_3-n),1+n\pm 2 \gamma_3\pm \gamma_4,(1\pm \gamma_3),\sin^2{\theta}\bigg]\bigg].
\end{eqnarray}
Let us now turn to the equation (\ref{EABB}). Putting the separation constant $A=(2n\pm \gamma_3\pm \gamma_4+1)^2$ into (\ref{EABB}), we have
\begin{eqnarray}
\bigg[-\frac{\partial^2}{\partial \rho^2}-\frac{1}{\rho} \frac{\partial}{\partial \rho}+c_1\rho^2+\frac{1}{\rho^2}\left(2n\pm \gamma_3\pm \gamma_4+1\right)^2+A_1\bigg]G(\rho) = 0. \label{EABBB}
\end{eqnarray}
By setting $\xi=\varepsilon \rho^2$ , $G(\xi)=\xi^{\alpha}G_1(\xi)$ and $G_1(\xi)=e^{-\frac{1}{2} \xi}G_2(\xi)$, (\ref{EABBB}) can be converted to the form
\begin{eqnarray}
\xi {G_2}^{''}(\xi)+\bigg[\left(2\alpha+1\right)-\xi \bigg]G_2^{'}(\xi)-\bigg[\frac{1}{2}\left(2\alpha+1\right)+\frac{A_1}{4\varepsilon}\bigg]G_2\left(\xi\right)=0,\label{MH1}
\end{eqnarray}
where $\alpha=\frac{1}{2}\left(1\pm\gamma_3\pm\gamma_4+2n\right)$ and $\varepsilon^2=c_1$.
Comparing (\ref{MH1}) with the confluent hypergeometric differential equation \cite{magn1},
\begin{eqnarray}
z f^{''}(z)+\left(c-z\right)f^{'}(z)-af(z)=0,
\end{eqnarray}
we obtain the separation constant,
\begin{eqnarray}
A_1=4\varepsilon a-2\varepsilon(2n\pm \gamma_3\pm \gamma_4+1)\label{AAA1}
\end{eqnarray}
and the solution of (\ref{EABBB}) can be written as
\begin{eqnarray}
&&G\left(\rho\right) =\bigg[\left(\varepsilon {\rho^2}\right)^{\frac{1}{2}\left(1\pm \gamma_3\pm \gamma_4+2n\right)} e^{-\frac{1}{2}\varepsilon \rho^2}  C_1 \phi\left(a+\frac{1}{2};\left(2\pm \gamma_3\pm \gamma_4+2n\right);\varepsilon\rho^2\right)\bigg]\nonumber\\&&\qquad\quad +\bigg[C_2 \left(\varepsilon\rho^2\right)^{1-\left(2\pm \gamma_3\pm \gamma_4+2n\right)} \phi \left(a+1;\left(2\pm \gamma_3\pm \gamma_4+2n\right);\varepsilon \rho^2\right)\bigg].
\end{eqnarray}
Putting (\ref{AAA1}) into (\ref{EABBBB}), we can be reformed as follows
\begin{eqnarray}
\bigg[\frac{\partial^2}{\partial z^2}-4 c_1 z^2-c_2 z+E+4 \varepsilon a- 2\varepsilon\left(2n\pm \gamma_3\pm \gamma_4+1\right)\bigg]F(z)=0,\label{EABBBBB}
\end{eqnarray}
which is one linear differential equation.
Such linear differential equation  
\begin{eqnarray}
f^{''}\left(z\right)+\left(-q^2z^2-2qsz+t\right)f\left(z\right)=0,\label{EABBBBBB}
\end{eqnarray}
solved in \cite{reh1} with the condition $q^{-1}\left(s^2+t\right)$ is an odd integer. Let us consider 
\begin{eqnarray}
q^{-1}\left(s^2+t\right)=2\tau+1, \quad \text{$\tau$ is an integer}.\label{odi1}
\end{eqnarray}
The elementary solution of (\ref{EABBBBBB}) is given \cite{reh1} as 
\begin{eqnarray}
F_1\left(z\right)=h\left(z\right). e^{-\frac{q}{2}\left(\left(z+\left(\frac{s}{q}\right)^2\right)\right)},
\end{eqnarray}
\begin{eqnarray}
F_2\left(z\right)=F_1\left(z\right).\int h\left(\xi\right)^{-2} e^{q\left(\xi+\frac{s}{q}\right)^2}d\xi,
\end{eqnarray}
where 
\begin{eqnarray}
h\left(z\right)=\left(z+\frac{s}{q}\right)^j+\sum_{1\le l \le \frac{j}{2}} \frac{j!}{l! \left(j-2l\right)! z^{2l}\left(-q\right)^l}\left(z+\frac{s}{q}\right)^{j-2l},\quad j=0,1,2, \dots.
\end{eqnarray}
Hence we obtain the solution of (\ref{EABBBBB}) as follows 
\begin{eqnarray}
F_1\left(z\right)=h\left(z\right). \,e^{-\sqrt{c_1}\left(\left(z+\frac{c_2}{4 c_1}\right)^2\right)}
\end{eqnarray}
\begin{eqnarray}
F_2\left(z\right)=F_1\left(z\right).\int h\left(\xi\right)^{-2} e^{2 \sqrt{c_1}\left(\xi+\frac{c_2}{4 c_1}\right)^2}d\xi
\end{eqnarray}
\begin{eqnarray}
h\left(z\right)=\left(z+\frac{c_2}{4 c_1}\right)^j+\sum_{1\le i \le \frac{j}{2}} \frac{j!}{i! \left(j-2i\right)! z^{2i}\left(-2\sqrt{c_1}\right)^i}\left(z+\frac{c_2}{4 c_1}\right)^{j-2i}.
\end{eqnarray}
Comparing (\ref{EABBBBB}) with (\ref{EABBBBBB}), (\ref{odi1}) and substituting $c_1=\gamma_1^2$, $c_2=\gamma_2$, we can obtain the eigenvalues of the nondegenerate system (\ref{Ham1}) in terms of quantum numbers involving the four parameters as follows,
\begin{eqnarray}
E=2(2\tau +1)\gamma_1+2(2n-2a\pm \gamma_3 \pm \gamma_4 +1)\gamma_1-\frac{\gamma_2^2}{16\gamma_1^2}.\label{Ecy1}
\end{eqnarray}
Making identification $2\tau=p_1+p_2+1$, $2(n-a)=n_1+n_1$, $2\gamma_3=m_3$ and $2\gamma_4=m_4$, the energy spectrum (\ref{Ecy1}) becomes (\ref{Eal1}).

\subsection{Paraboloidal coordinates}
The paraboloidal coordinates are considered by 
\begin{eqnarray}
x_1=\frac{1}{2} (u^2-v^2),\qquad x_2= u v\sin{\phi} ,\qquad x_3= u v \cos{\phi},
\end{eqnarray}
where $0\le\phi<2\pi$, $u\ge0$ and $ v\ge0$.
Now the Schr\"{o}dinger eigenvalue equation $H\psi = E\psi$ of the system (\ref{Ham1}) in these coordinates leads to the following structure, 
\begin{eqnarray}
&& \bigg[-\bigg[\frac{\partial^2}{\partial\phi^2}+\frac{u^2v^2}{u^2+v^2}\left(\frac{\partial^2}{\partial u^2}+\frac{\partial^2}{\partial v^2}\right)+\frac{u^2v^2}{u^2+v^2}\left(\frac{1}{u}\frac{\partial}{\partial u}+\frac{1}{v}\frac{\partial}{\partial v}\right)\bigg]+\frac{c_2}{2} u^2 v^2\left(u^2-v^2\right) \nonumber\\&&\qquad +c_1 u^2 v^2\left(u^4+v^4-u^2 v^2\right) +\frac{c_3}{\sin^2{\phi}}+\frac{c_4}{\cos^2{\phi}}\bigg]\psi(u,v,\phi)=0.\label{EC}
\end{eqnarray}
To separate the Schr\"{o}dinger equation (\ref{EC}), the ansatz 
\begin{eqnarray}
\psi(u,v,\phi)=R(u,v)\,  Y(\phi)
\end{eqnarray}
gives rise to the following differential equations,
\begin{eqnarray}
&&\bigg[- \frac{\partial^2}{\partial \phi^2}+\frac{c_3}{ \sin^2\phi}+\frac{c_4}{\cos^2\phi}-A\bigg]Y(\phi)=0,\label{ECC}
\\&&
\bigg[-\frac{u^2v^2}{u^2+v^2}\left(\frac{\partial^2}{\partial u^2}+\frac{\partial^2}{\partial v^2}\right)-\frac{u^2v^2}{u^2+v^2}\left(\frac{1}{u}\frac{\partial}{\partial u}+\frac{1}{v}\frac{\partial}{\partial v}\right)\nonumber\\&& +c_1 u^2 v^2\left(u^4+v^4-u^2 v^2\right)+\frac{c_2}{2} u^2 v^2\left(u^2-v^2\right) -E u^2v^2+A\bigg]R(u,v)=0,\label{ECCC}
\end{eqnarray}
where $A$ is a separation constant. Again taking the ansatz
\begin{eqnarray}
R(u,v)=R_1 (u) R_2 (v)
\end{eqnarray}
for the separation of (\ref{ECCC}), it leads to
\begin{eqnarray}
&&\bigg[-\frac{\partial^2}{\partial u^2}-\frac{1}{u}\frac{\partial}{\partial u}+c_1 u^6+\frac{c_2}{2}u^4-Eu^2+\frac{A}{u^2}-A_1\bigg]R_1 (u)=0,\label{ECCCC}
\\&&
\bigg[-\frac{\partial^2}{\partial v^2}-\frac{1}{v}\frac{\partial}{\partial v}+c_1 v^6+\frac{c_2}{2}v^4-Ev^2+\frac{A}{v^2}+A_1\bigg]R_2 (v)=0.\label{ECCCCC}
\end{eqnarray}
We now first change to (\ref{ECC}) by setting $w=\sin^2{\phi}$ and $g(w)=w^{\alpha_1} (1-w)^{\alpha_2}{g_1}(w) $, which can be reduced to the following form 
\begin{eqnarray}
 w(1-w) {g_1}^{''}+\bigg[\left(1\pm {\gamma}_3\right)-\left(2\pm \gamma_3\pm \gamma_4\right)w\bigg]{g_1}^{'} -\bigg[\left(\frac{1}{2}\pm \frac{\gamma_3}{2}\pm \frac{\gamma_4}{2}\right)^2- \frac{A}{4} \bigg]g_1(w)=0,\label{ECCCCCC}
\end{eqnarray}
 where $ \gamma_3=\pm \frac{1}{2} \sqrt{1+4 c_3}$ and  $ \gamma_4=\pm \frac{1}{2} \sqrt{1+4 c_4}$ , $\alpha_1=\frac{1}{4}\pm\frac{\gamma_3}{2}$ and $\alpha_2=\frac{1}{4}\pm\frac{\gamma_4}{2}$.
Comparing (\ref{ECCCCCC}) in terms of the Jacobi differential equation \cite{magn1},  
\begin{eqnarray}
\mathcal{X}\left(1-\mathcal{X}\right)\mathcal{Y}^{''}+\left[\gamma-\left(\alpha+1\right)\mathcal{X}\right]\mathcal{Y}^{'}+\eta\left(\alpha+\eta\right)\mathcal{Y}=0\label{ECCCCCCC}
\end{eqnarray}
and its solution,
\begin{eqnarray}
&& y=C_1 {}_2F_1\left(-\eta, \eta+\alpha, \gamma, \mathcal{X}\right)\nonumber\\&& \qquad -\left(-1\right) \mathcal{X}^{1-\gamma} C_{2} {}_2F_1 \left(1-\eta-\gamma, 1+\eta +\alpha-\gamma, 2-\gamma, \mathcal{X}\right),
\end{eqnarray}
we obtain the separation constant 
\begin{eqnarray}
A=\left(2\eta\pm \gamma_3 \pm \gamma_4+1\right)^2,
\end{eqnarray}
and the solution of (\ref{ECC}), 
\begin{eqnarray}
&& Y\left(\phi\right)=({\sin^2{\phi}})^{\frac{1}{4}\pm \frac{\gamma_3}{2}} (1-\sin^2{\phi})^{\frac{1}{4}\pm \frac{\gamma_4}{2}}  \bigg[C_{1\,2} F_1\left(-\eta, \eta+1\pm \gamma_3\pm \gamma_4,1\pm \gamma_3,\sin^2{\phi}\right)\nonumber\\&& -\left(-1\right)^{-(1\pm \gamma_3)} \left(\sin^2 \phi\right)^{\pm {\gamma_3}}  C_{2\,2}F_1\bigg[(\pm \gamma_3-\eta),1+\eta\pm 2 \gamma_3\pm \gamma_4,(1\pm \gamma_3),\sin^2{\phi}\bigg]\bigg].
\end{eqnarray}
To solve the differential equations (\ref{ECCCC}) and (\ref{ECCCCC}), let us set $z_1=u^2$ in (\ref{ECCCC}) and $z_2=v^2$ in (\ref{ECCCCC}), the the couple equations become,
\begin{eqnarray}
\bigg[z_i^2\frac{\partial^2}{\partial z_i^2}+z_i\frac{\partial}{\partial z_i}+\left( -\frac{c_1}{4}z_i^4-\frac{c_2}{8}z_i^3+\frac{E}{4}z_i^2-\frac{A}{4}+\frac{A_i}{4}z_i\right)\bigg]R_i\left(z_i\right)=0,\label{ECCA}
\end{eqnarray} 
where $A_1=-A_2$ and $i=1, 2$.
The equation (\ref{ECCA}) can be transformed into a Bi-Confluent Heun differential equation \cite{Ronv1,Ish1} of type
\begin{eqnarray}
\mathcal{X}\, F^{''}+\left(1+p-q\mathcal{X}-2 \mathcal{X}^2\right) \, F^{'}+\bigg[\left(r-p-2\right)\mathcal{X}-\frac{1}{2}\left(s+q\left(1+p\right)\right)\bigg]F\left(\mathcal{X}\right)=0,\label{ECCD}
\end{eqnarray}
which has a solution in terms of Hermite functions,
\begin{eqnarray}
F=\sum_{n=0}^\infty\,c_n  H_{n+1+p+\frac{1}{2}\left(r-p-2\right)}\left(\mathcal{X}+\frac{q}{2}\right),
\end{eqnarray}
where $c_n$ satisfies the three terms recurrence formulas,
\begin{eqnarray}
c_n L_n +c_{n-1} Q_{n-1} + c_{n-2} P_{n-2} =0.
\end{eqnarray}
with the relations
\begin{eqnarray}
&& L_n=2n\left(p+n+\frac{\left(r-p-2\right)}{2}+1\right), \qquad P_n=p+n+1, \nonumber\\ && Q_n=-\frac{1}{2}\left(s+q\left(p+1\right)\right)+q\left(p+n+1\right).
\end{eqnarray}
By setting $R_i\left(z_i\right)=z_i^{\rho}\, e^{az_i+\frac{b}{2}z_i^2}\, f_i\left(y\right)$ and $z_i=ky_i$ into (\ref{ECCA}), it leads to
\begin{eqnarray}
&& y_i^2{ f_i}^{''}(y_i)+y_i\left(1+2\rho+ 2 a k y_i +2bk^2 y_i^2\right){f_i}^{'}(y_i) +\bigg[\left(\rho^2 -\frac{A}{4}\right)+\left(2a\rho+a+\frac{A_1}{4}\right)ky_i \nonumber\\&& +\left(2b\rho+2b+a^2+\frac{E}{4}\right)k^2 y_i^2 +\left(2ab-\frac{c_2}{8}\right)k^3 y_i^3+\left(b^2 -\frac{c_1}{4}\right)k^4 y_i^4\bigg]f_i\left(y_i\right)=0.\label{ECCF}
\end{eqnarray}
To compare (\ref{ECCH}) and (\ref{ECCD}) with the suitable choice of signs, we have the following conditions, 
\begin{eqnarray}
\rho=\sqrt{\frac{A}{4}},\qquad b=-\frac{\sqrt{c_1}}{2}, \qquad a=-\frac{c_2}{8\sqrt{c_1}},\qquad k^2=\frac{2}{\sqrt{c_1}},\label{ECCG}
\end{eqnarray}
Using the above conditions, we can rewrite (\ref{ECCF}) as follows 
\begin{eqnarray}
&& y_i{f_i}^{''}(y_i)+\left(1+2\sqrt{\frac{A}{4}}-\frac{c_2}{2c_1}y_i-2y_i^2\right){f_i}^{'}(y_i) +\bigg[ \left(-\frac{c_2\, A}{16\sqrt{c_1}}-\frac{c_2}{8\sqrt{c_1}}+\frac{A_1}{4}\right)\left(\sqrt{\frac{2}{\sqrt{c_1}}}\right)\nonumber\\&&\qquad + \left(-\sqrt{\frac{c_1\, A}{4}}-\sqrt{c_1}+\frac{c_2^2}{64 c_1}+\frac{E}{4}\right) \left(\frac{2}{\sqrt{c_1}}\right)y_i\bigg]f_i\left(y_i\right)=0\label{ECCH}
\end{eqnarray}
Again by comparing (\ref{ECCH}) and (\ref{ECCD}), we obtain
\begin{eqnarray}
r-p-2=\left(-\sqrt{\frac{c_1\, A}{4}}-\sqrt{c_1}+\frac{c_2^2}{64 c_1}+\frac{E}{4}\right) \left(\frac{2}{\sqrt{c_1}}\right)
\end{eqnarray}
One can shown that the solution of a Bi-Confluent Heun equation \cite{Ronv1,Ish1} is the $n$ degree polynomial, then it allows the condition $r-p-2=2\mu$, $\mu$ is an integer. Using this condition,
the solution of (\ref{ECCCC}) is given as
\begin{eqnarray}
f_i\left(y_i\right)=y_i^{\sqrt{A}}\, e^{\left({\frac{c_2}{8\sqrt{c_1}}y_i^2}+\frac{\sqrt{c_1}}{4}y_i^4\right)}\sum_{n=0}^\infty\,c_n  H_{n+1+p+\frac{1}{2}\left(r-p-2\right)}\left(y_i^2+\frac{q}{2}\right),
\end{eqnarray} 
where
\begin{eqnarray}
&& p=\sqrt{A},\quad r=\frac{c_2^2}{32 c_1^{\frac{3}{2}}}+\frac{E}{2\sqrt{c_1}},\quad q=\frac{c_2}{2 c_1},\quad A=\left(2\eta\pm\gamma_3\pm\gamma_4+1\right)^2.
\end{eqnarray}
Then we can present the spectrum explicit relation,
\begin{eqnarray}
E=\sqrt{c_1}\left(4+4\mu\right)+2\sqrt{c_1\, A}-\frac{c_2^2}{16 c_1}.\label{ECCI}
\end{eqnarray}
Substituting $c_1=\gamma_1^2$, $c_2=\gamma_2$ and $A=\left(2\eta\pm \gamma_3\pm \gamma_4+1\right)^2$ into (\ref{ECCI}), we have the required eigenvalue of the system (\ref{Ham1}),
\begin{eqnarray}
E= 4(\mu+1)\gamma_1 +2(2\eta\pm \gamma_3 \pm \gamma_4+1)\gamma_1 -\frac{\gamma_2^2}{16 \gamma_1^2}.\label{ECCJ}
\end{eqnarray}
Making identification $2\mu=p_1+p_2+1$, and $2\eta=n_1+n_1$, the energy spectrum (\ref{ECCJ}) becomes (\ref{Eal1}).

\section{Conclusions} \label{Se5}
We constructed the quadratic full symmetry algebra for the 3D nondegenerate quantum superintegrable system generated by six linearly independent integrals of motion including the Hamiltonian. The symmetry algebra contains quadratic subalgebra structures generated by three generators with structure constants connected to the quadratic algebra of the two-dimensional quantum superintegrable system \cite{das01}. The algebraic calculations of the symmetry algebra to the quantum superintegrable system enable us to obtain the energy spectrum. We have presented corresponding Casimir invariants and derived the structure functions of the quadratic subalgebras of the symmetry algebra in the realizations of deformed oscillators. The finite-dimensional unirreps of these structure functions yield the energy spectrum of the model algebraically. We also showed that the system is multiseparable in cylindrical polar and paraboloidal coordinates. We solved the Schr\"{o}dinger equation of the system and expressed the wave functions in terms of special functions, and obtained the physical spectrum. The results are compared with those spectrum obtained from the algebraic computation.
\\[3mm]

\section*{Acknowledgements}
\noindent
MA was supported by the National Science and Technology Fellowship, Bangladesh. FH was partially supported by the project grant CZ. 02.2.69/0.0/0.0/18 \_0016980, co-financed by the European Union. The authors thank Ian Marquette (University of Queensland, Australia) and Libor Snobl (Czech Technical University in Prague, Czech Republic) for constructive discussions and helpful comments on the subject of this manuscript.

\end{document}